# Magnetic and transport properties in magnetic topological insulators MnBi$_2$Te$_4$(Bi$_2$Te$_3$)$_n$ (n=1,2)

M.Z. Shi[1], B. Lei[1], C. S. Zhu[1], D. H. Ma[1], J. H. Cui[1], Z.L. Sun[1], J. J. Ying[1,2,*], X. H. Chen[1,2,3,†]

1. Hefei National Laboratory for Physical Sciences at the Microscale and Department of Physics, and Key Laboratory of Strongly-coupled Quantum Matter Physics, Chinese Academy of Sciences (CAS), University of Science and Technology of China, Hefei, Anhui 230026, China

2. CAS Center for Excellence in Quantum Information and Quantum Physics, Hefei, Anhui 230026, China

3. Collaborative Innovation Center of Advanced Microstructures, Nanjing University, Nanjing 210093, China

*E-mail: yingjj@ustc.edu.cn
†E-mail: chenxh@ustc.edu.cn

**The observation of quantized anomalous Hall conductance in the forced ferromagnetic state of MnBi$_2$Te$_4$ thin flakes has attracted much attentions. However, strong magnetic field is needed to fully polarize the magnetic moments due to the large antiferromagnetic interlayer exchange coupling. Here, we reported the magnetic and electrical transport properties of the magnetic van der Waals MnBi$_2$Te$_4$(Bi$_2$Te$_3$)$_n$ (n=1,2) single crystals, in which the interlayer antiferromagnetic exchange coupling is greatly suppressed with the increase of the separation layers Bi$_2$Te$_3$. MnBi$_4$Te$_7$ and MnBi$_6$Te$_{10}$ show weak antiferromagnetic transition at 12.3 and 10.5 K, respectively. The ferromagnetic hysteresis was observed at low temperature for both of the crystals, which is quite crucial for realizing the quantum anomalous Hall effect without external magnetic field. Our work indicates that MnBi$_2$Te$_4$(Bi$_2$Te$_3$)$_n$ (n=1,2) provide ideal platforms to investigate the rich topological phases with going to their 2D limits.**

## I. INTRODUCTION

The quantum anomalous Hall (QAH) effect which is induced by spontaneous magnetization without any external magnetic field has attracted much attentions for its potential applications in the electronic and spintronic devices due to the dissipationless spin-polarized chiral edge states [1-11]. QAH was firstly realized in Cr- or V-doped (Bi,Sb)$_2$Te$_3$ magnetic topological thin films after it had been sought for over two decades [3,4,12]. However, the extremely low working temperature (usually <100 mK) impedes its further applications. In addition, the unavoidable randomly distributed magnetic impurities induce strong inhomogeneity in the electronic structure and magnetic properties [13]. Thus, the intrinsic magnetic topological insulators with homogeneous electronic and magnetic properties are highly desired, and it is expected that the QAH working temperature is remarkably increased.

MnBi$_2$Te$_4$ is a stoichiometric tetradymite-type compound with Te-Bi-Te-Mn-Te-Bi-Te septuple layers (SLs) stacking along the c-axis. Each SL is formed from the intercalation of a MnTe bilayer into a quintuple layer (QL) of $Bi_2Te_3$. SLs are coupled through van der Waals bonding, thus this material can be thinned down to 2D atomic thin layers through microexfoliation. $MnBi_2Te_4$ was suggested as the first example of antiferromagnetic topological insulator (AFM TI) [14-19]. $MnBi_2Te_4$ was determined as the A-type AFM with the transition temperature around 25 K. By applying high magnetic field, AFM state can be tuned to fully polarized ferromagnetism (FM) [20-22]. Recently, quantized anomalous Hall conductance was observed in the $MnBi_2Te_4$ thin flakes after the magnetic moments were fully polarized along *c*-axis direction at much higher temperature (4.5 K) [23,24]. However, a large external magnetic field (>6 T) is needed to fully polarize the AFM spins into a forced FM state. In order to realize the quantized anomalous Hall conductance at much lower magnetic field or even zero field, interlayer antiferromagnetic exchange coupling needs to be greatly weakened. One method is by intercalating nonmagnetic separation, such as the recently discovered $MnBi_2Te_4(Bi_2Te_3)_n$ (n=1,2)[25].

$MnBi_2Te_4(Bi_2Te_3)_n$ (n=1,2) are identified as natural van der Waals heterostructure in which SLs are separated by one or two $Bi_2Te_3$ QLs, respectively. Since the magnetic $MnBi_2Te_4$ SLs are far separated by the nonmagnetic $Bi_2Te_3$ QLs, the interlayer antiferromagnetic exchange coupling is expected to be greatly weakened. The recent experiments indeed indicate the much weaker interlayer antiferromagnetic coupling in $MnBi_4Te_7$ and the topological electronic structure was observed [26-28]. In the 2D limit of $MnBi_4Te_7$ and $MnBi_6Te_{10}$, different stacking sequences of the $MnBi_2Te_4$ SLs and $Bi_2Te_3$ QLs can be formed, and can lead to various topological phases including QAH [29]. Dirac surface states have been observed in similar compounds $PbBi_4Te_7$ and $PbBi_6Te_{10}$ indicating the possible topological properties in this type of materials[30]. Thus, both of $MnBi_4Te_7$ and $MnBi_6Te_{10}$ crystals provide an ideal platform to study the topological phase transition and to realize the QAH effect. However, due to the difficulty of growing single crystals, detailed physical property measurements for both of the crystals are still lacking. Here, we successfully synthesized $MnBi_2Te_4(Bi_2Te_3)_n$ (n=1,2) single crystals by using the self-flux method, and studied the magnetic and electrical transport properties of the single crystals in detail. Both of the single crystals show weak AFM transition at 12.3 K and 10.5 K, respectively. Our work demonstrates that the interlayer AFM exchange coupling is greatly weakened by increasing the $Bi_2Te_3$ inserting layers in $MnBi_2Te_4(Bi_2Te_3)_n$ (n=1,2) crystals. At low temperature, both of the crystals show FM hysteresis, and the forced FM state can be stabilized even at zero field in these crystals, which is quite crucial for realizing QAH.

## II. METHODS

The $MnBi_2Te_4(Bi_2Te_3)_n$ (n=1,2) single crystals were synthesized through a self-flux method. For the synthesis of $MnBi_4Te_7$ single crystal, the Mn powder (99.95%), Bi lump (5N) and Te lump(5N) were weighted and mixed with the molar ratio of Mn:Bi:Te = 1:4:7. The mixture was sealed into a carbon coated silica container, and reacted in a self-built vertical Bridgman furnace. The furnace was heated to 1000 °C in 500 minutes, and kept for 24 hrs; then the furnace was cooled to 720 °C in 24 hrs, and further cooled to 600 °C in 48 hrs and kept for another 24 hrs. Finally, the ampoule was quenched in water quickly. For the synthesis of $MnBi_6Te_{10}$ single crystal, the molar ratio of Mn powder, Bi lump and Te lump was changed to Mn:Bi:Te=1:6:10. The same temperature control process as that of $MnBi_4Te_7$ was adopted to obtain the $MnBi_6Te_{10}$ single crystal.

All the single crystals were checked using X-ray diffractometer (SmartLab-9, Rikagu Corp.) with Cu K$_\alpha$ radiation. The magnetization data were obtained using a SVSM magnetometer, and the electric transport properties were measured with the Quantum Design PPMS-9.

## III. RESULTS AND DISCUSSION

The XRD patterns (Figure 1a and 1b) of the obtained crystals show the sharp (00L) diffraction peaks, and are well consistent with the previous reports [25-27], indicating the pure phase of MnBi$_4$Te$_7$ and MnBi$_6$Te$_{10}$ with a [001] preferred orientation direction. The XRD patterns of MnBi$_4$Te$_7$ and MnBi$_6$Te$_{10}$ with Y-axis in log-scale (see Fig.S1 in supplementary materials) further confirm the high purity and quality of the single crystals. The calculated lattice parameter of c-axis can be determined to be 23.781 Å and 101.774 Å for MnBi$_4$Te$_7$ and MnBi$_6$Te$_{10}$, respectively. The space group of MnBi$_6$Te$_{10}$ is R-3m[31], thus the c-axis lattice constant is treble the distance of the adjacent MnTe layer. The distance of the adjacent MnTe layer in MnBi$_4$Te$_7$ and MnBi$_6$Te$_{10}$ are 23.781 Å and 33.925 Å, respectively. The proposed crystal structure models of MnBi$_4$Te$_7$ and MnBi$_6$Te$_{10}$ are shown in Figure 1c and 1d[31], respectively. The structure of MnBi$_4$Te$_7$ can be regarded as the van der Waals heterostructure in which the MnBi$_2$Te$_4$ monolayer is spaced with monolayer Bi$_2$Te$_3$, while MnBi$_6$Te$_{10}$ can be regarded as the van der Waals heterostructure which consists of monolayer MnBi$_2$Te$_4$ and double layers Bi$_2$Te$_3$. As a result, the distance of the adjacent magnetic MnTe layer gradually increases. It leads to a decrease in the interlayer AFM coupling and induces competition between different magnetic ground state(for example FM and AFM). The magnetic susceptibility of MnBi$_4$Te$_7$ under a small magnetic field (50 Oe) applied along c-axis and in ab-plane is shown in Fig. 1e, indicating that an AFM transition takes place at around 12.3 K. The zero field-cooled (ZFC) and field-cooled (FC) magnetic susceptibility shows a discrepancy with external magnetic field along c-axis. We can fit the high-temperature susceptibility measured at 0.5 Tesla to the Curie-Weiss law as shown in the inset of Fig. 1e. The parameter $\chi_0$ is the temperature independent term which contains the core diamagnetism and Van Vleck paramagnetism. The effective moments are $\mu_{eff}^{ab}$=5.0 $\mu_B$/Mn and $\mu_{eff}^{c}$=4.9 $\mu_B$/Mn, and Weiss temperatures are $\Theta_w^{ab}$=11.4 K and $\Theta_w^{c}$ = 12.3 K, in consistent with previous reports[26,27]. The ZFC and FC magnetic susceptibility for MnBi$_6$Te$_{10}$ show similar behavior to that of MnBi$_4$Te$_7$ as shown in Fig. 1f with AFM transition at 10.5 K. By fitting the high-temperature susceptibility measured at 0.5 T with the Curie-Weiss law, we can extract the effective moments are $\mu_{eff}^{ab}$=4.8 $\mu_B$/Mn, $\mu_{eff}^{c}$=4.7 $\mu_B$/Mn, and Weiss temperatures are $\Theta_w^{ab}$= 9.9 K and $\Theta_w^{c}$= 13.0 K. The small hump around 7 K in the ZFC curve is possibly due to the slight Mn occupied at Bi sites in the quintuple layer of Bi$_2$Te$_3$. As shown in Fig.1g, temperature dependences of resistivity for MnBi$_4$Te$_7$ and MnBi$_6$Te$_{10}$ single crystals show an anomaly at around 12.3 K and 10.5 K, respectively, which are consistent with the results of magnetic susceptibility. Both of crystals show metallic behavior similar to the MnBi$_2$Te$_4$ single crystal [20-22]. The anomalies around magnetic transition temperatures can be gradually suppressed with increasing the applied magnetic field (see Fig. S2 in the supplementary materials). Such observation is consistent with the spin-fluctuation-driven spin scattering scenario [22]. In these crystals, spin scattering generated by interlayer AFM coupling results in a high resistivity state, while strong external field can suppress it and tune the system to a forced FM state [32,33].

These behaviors are similar to the case of MnBi$_2$Te$_4$, indicating a similar magnetic structure in MnBi$_4$Te$_7$ and MnBi$_6$Te$_{10}$.

In order to further investigate the magnetic properties of MnBi$_2$Te$_4$(Bi$_2$Te$_3$)$_n$ (n=1,2) single crystals, we performed the field dependent magnetization measurements at various temperatures. Figure 2a and 2b show the isothermal magnetization for MnBi$_4$Te$_7$ under magnetic field applied along c-axis and in ab-plane at various temperatures, respectively. The magnetic easy axis is along c-axis direction because much smaller magnetic field is needed to fully polarize the spins with H//c, being the same as that in MnBi$_2$Te$_4$. The magnetic moment can be easily aligned in this system due to the weak interlayer AFM exchange coupling. We observed a spin-flip transition and small magnetic field (less than 0.2 T) can fully polarize the magnetic moments with the field along c-axis direction as shown in Fig. 2c. While the magnetic moments can be gradually polarized by the external magnetic field applied in the ab-plane. These *M-H* results indicate the magnetic moments are aligned antiferromagnetically along c-axis direction, similar to the case of MnBi$_2$Te$_4$. Surprisingly, ferromagnetic hysteresis emerges at low temperature as shown in Fig. 2c. The emergence of low temperature ferromagnetic hysteresis can be well explained by the energy barrier between the AFM and forced FM states caused by the magnetocrystalline anisotropy. The interlayer AFM exchange coupling is greatly weakened, and cannot overcome the energy of magnetocrystalline anisotropy in this system, so that a FM hysteresis emerges at low temperature. At higher temperature, the thermal fluctuation overcomes the energy barrier, and no FM hysteresis can be observed. We can estimate the magnetocrystalline anisotropy energy by using the saturation field of the *M-H* curves. We used difference of anisotropy saturation field $\mu_0 \Delta H=1$ T and effective moments as 5 $\mu_B$/Mn, thus the anisotropy energy barrier can be estimated to be 0.29 meV/Mn which is comparable with the thermal energy (5 K) [27]and the interlayer exchange coupling (-0.15 meV/Mn from the calculation)[26]. We should point out that there are strong competition between FM and AFM exchange couplings in MnBi$_4$Te$_7$ because the coercive field does not monotonously increase with decreasing the temperature, and FM hysteresis also appears at 2 K with H//ab (see Fig.S5a in supplementary materials).

Figure 2d and 2e show the isothermal magnetization at various temperatures for MnBi$_6$Te$_{10}$ with magnetic field applied along c-axis and in ab-plane, respectively. Similar to MnBi$_4$Te$_7$, much higher magnetic field is needed to fully polarize the spins with H//ab, indicating that magnetic easy axis is along c-axis direction. As shown in Fig. 2f, the *M-H* curve at 2 K also shows FM hysteresis at low temperature, similar to MnBi$_4$Te$_7$. At 2 K, by increasing the magnetic field along c-axis direction, the spin-flip transition happens and the magnetic moments are fully polarized at 0.2 Tesla. When the field is decreased, the forced FM state can be stabilized even at zero field and small negative magnetic field (0.02 Tesla) is needed to tune the forced FM to AFM due to the magnetocrystalline anisotropy. With increasing the temperature, the FM hysteresis gradually disappears due to the thermal fluctuation. The *M-H* curves for MnBi$_6$Te$_{10}$ are similar to that of MnBi$_4$Te$_7$, indicating the same magnetic structure as that of MnBi$_4$Te$_7$. Very small field (less than 0.1 Tesla) can fully polarize the magnetic moments above 4 K, indicating a weaker interlayer AFM exchange coupling relative to MnBi$_4$Te$_7$.

In order to further understand the magnetic properties due to the spin-charge interaction, we performed the isothermal magnetoresistance measurements on MnBi$_4$Te$_7$ and MnBi$_6$Te$_{10}$ single crystals as shown in Fig. 3. For MnBi$_4$Te$_7$, the behavior of $\rho_{xx}$–H is closely related to the AFM-FM spin-flip transition driven by magnetic field applied along c-axis. As shown in Fig. 3a, the butterfly shape magnetoresistance at 2 K is consistent with the FM hysteresis at low temperature in magnetization measurements [26,27]. With H//ab, the in-plane resistivity $\rho_{xx}$ continuously decreases below 2 Tesla due to the gradually polarization of the magnetic moments as shown in Fig. 3b. The behavior of $\rho_{xx}$ for the MnBi$_6$Te$_{10}$ crystal is quite similar to that of MnBi$_4$Te$_7$ as shown in Fig. 3c and 3d with magnetic field applied along c-axis and in ab-plane, respectively. The sharp change of the magnetoresistance is related to the spin-flip transition with magnetic field applied along c-axis direction. Compared to MnBi$_2$Te$_4$, the magnetoresistance of MnBi$_4$Te$_7$ shows an obvious upturn at high magnetic field(see Fig. S3a in supplementary materials), and MnBi$_6$Te$_{10}$ exhibits large positive magnetoresistance (see Fig. S3b in supplementary materials) at high magnetic field. Such large positive magnetoresistance should origin from the contribution of Bi$_2$Te$_3$ QLs.

We also performed the Hall resistivity measurements on the MnBi$_2$Te$_4$(Bi$_2$Te$_3$)$_n$ (n=1,2) single crystals as shown in Figure 4 with field applied along c-axis direction. The carriers for both of the crystals are electron-type, similar to the case of MnBi$_2$Te$_4$. We can clearly observe the anomalous Hall effect in the magnetic ordering state of MnBi$_4$Te$_7$ as shown in Fig. 4b. The anomalies in the Hall resistivity are due to the spin-flip transitions, consistent with the M-H and magnetoresistance measurements. However, the anomalous Hall effect for MnBi$_6$Te$_{10}$ is very weak (see Fig. S4 in the supplementary materials). We can obtain the anomalous Hall resistivity $\rho_{xy}^A$ of MnBi$_6$Te$_{10}$ by subtracting the linear Hall term as shown in Fig. 3c. The anomalies in the $\rho_{xy}^A$ are due to the spin-flip transition, which is consistent with the observations in M-H and magnetoresistance measurements. We can calculate anomalous Hall conductance $\sigma_{xy}^A$ ( $\approx \rho_{xy}^A/\rho_{xx}^2$) as 12.2 and 1.3 $\Omega^{-1}$cm$^{-1}$ for MnBi$_4$Te$_7$ and MnBi$_6$Te$_{10}$, respectively. Thus, the anomalous Hall conductivity per magnetic layer for MnBi$_4$Te$_7$ and MnBi$_6$Te$_{10}$ can be estimated to be 0.075 and 0.011 e$^2$/h, respectively. These values are comparable with the MnBi$_2$Te$_4$ crystals (0.063 e$^2$/h) reported previously[34]. Even though the anomalous Hall conductivity of bulk MnBi$_2$Te$_4$, MnBi$_4$Te$_7$ and MnBi$_6$Te$_{10}$ is only a few percent of the von Klitzing constant, it is still possible for the thin flake sample of MnBi$_4$Te$_7$ and MnBi$_6$Te$_{10}$ to realize QAH effect just like that of MnBi$_2$Te$_4$. The different anomalous Hall conductivity may possibly be related to the different electronic structure. We can obtain temperature dependent carrier density calculated from the linear $\rho_{xy}$(H) term at high magnetic field for MnBi$_4$Te$_7$ and MnBi$_6$Te$_{10}$. Temperature dependence of the carrier density for both of the crystals shows a weak temperature dependence as shown in Fig. 4d, similar to the behavior observed in normal metal. The carrier density for both MnBi$_4$Te$_7$ and MnBi$_6$Te$_{10}$ crystals is around 3×10$^{20}$ cm$^{-3}$.

We demonstrated that the interlayer antiferromagnetic exchange coupling is greatly weakened by inserting the nonmagnetic Bi$_2$Te$_3$ QLs into MnBi$_2$Te$_4$ SLs in MnBi$_2$Te$_4$(Bi$_2$Te$_3$)$_n$ (n=1,2). The AFM transition temperature is suppressed to 12.3 K and 10.5 K for MnBi$_2$Te$_4$(Bi$_2$Te$_3$) and MnBi$_2$Te$_4$(Bi$_2$Te$_3$)$_2$, respectively. Spin-flip transition occurs in both of the crystals at a small magnetic field of 0.1 ~ 0.2 Tesla applied along c-axis direction at low temperature, indicating that the magnetic structure for both of the crystals is A-type AFM, similar to the case of

$MnBi_2Te_4$. A FM hysteresis appears at low temperature for both of the crystals due to the weak interlayer AFM exchange coupling and magnetocrystalline anisotropy, and such FM hysteresis demonstrates that the fully polarized FM state can be stabilized even at zero field. It is quite crucial for realizing the QAH at their 2D limit. In addition, various stacking sequences of $MnBi_2Te_4$ SLs and $Bi_2Te_3$ QLs can be achieved for $MnBi_4Te_7$ and $MnBi_6Te_{10}$ single crystals through microexfoliation. The DFT calculation predicts that different stacking sequences lead to different topological states [29], so that $MnBi_2Te_4(Bi_2Te_3)_n$ (n=1,2) provide an ideal platform to study the fertile topological phases by flexibly tuning the stacking sequence. Further experiments on thin flakes are highly desired to realize the predicted topological phases in $MnBi_2Te_4(Bi_2Te_3)_n$ (n=1,2) thin flakes. The topological surface states are strongly correlated with the magnetic structures in these crystals, our work will help people to study the topological electronic structures in the future.

## IV. CONCLUSION

In conclusion, we systematically studied the magnetic properties of magnetic topological insulators $MnBi_2Te_4(Bi_2Te_3)_n$ (n=1,2) with van der Waals coupling. The interlayer AFM exchange couplings are greatly weakened due to the increase of $Bi_2Te_3$ layers in $MnBi_2Te_4(Bi_2Te_3)_n$, and the weakening of the interlayer AFM exchange couplings leads to the decrease of AFM transition temperature relative to $MnBi_2Te_4$. $MnBi_2Te_4(Bi_2Te_3)_n$ (n=1,2) provide us ideal platforms to investigate the topological properties.

**Acknowledgements**

This work is supported by the National Key R&D Program of China (Grant Nos. 2016YFA0300201 and 2017YFA0303001), the National Natural Science Foundation of China (Grant Nos. 11534010 and 11888101), the Key Research Program of Frontier Sciences, CAS, China (QYZDY-SSW-SLH021), the Strategic Priority Research Program (B) of the Chinese Academy of Sciences (Grant No. XDB25010100), Science Challenge Project (Grants No. TZ2016004), and Hefei Science Center CAS (2016HSC-IU001).

References:

[1] C.-X. Liu, S.-C. Zhang, and X.-L. Qi, Annual Review of Condensed Matter Physics **7**, 301 (2016).
[2] C.-Z. Chang and M. Li, Journal of Physics: Condensed Matter **28**, 123002 (2016).
[3] C.-Z. Chang, W. Zhao, D. Y. Kim, H. Zhang, B. A. Assaf, D. Heiman, S.-C. Zhang, C. Liu, M. H. W. Chan, and J. S. Moodera, Nature Materials **14**, 473 (2015).
[4] C.-Z. Chang, J. Zhang, X. Feng, J. Shen, Z. Zhang, M. Guo, K. Li, Y. Ou, P. Wei, L.-L. Wang, Z.-Q. Ji, Y. Feng, S. Ji, X. Chen, J. Jia, X. Dai, Z. Fang, S.-C. Zhang, K. He, Y. Wang, L. Lu, X.-C. Ma, and Q.-K. Xue, Science **340**, 167 (2013).
[5] R. Yu, W. Zhang, H.-J. Zhang, S.-C. Zhang, X. Dai, and Z. Fang, Science **329**, 61 (2010).
[6] C.-X. Liu, X.-L. Qi, X. Dai, Z. Fang, and S.-C. Zhang, Physical Review Letters **101**, 146802 (2008).
[7] H. Weng, R. Yu, X. Hu, X. Dai, and Z. Fang, Advances in Physics **64**, 227 (2015).
[8] X. Kou, Y. Fan, M. Lang, P. Upadhyaya, and K. L. Wang, Solid State Communications **215-216**, 34 (2015).

[9] Q.-Z. Wang, X. Liu, H.-J. Zhang, N. Samarth, S.-C. Zhang, and C.-X. Liu, Physical Review Letters **113**, 147201 (2014).
[10] K. He, Y. Wang, and Q.-K. Xue, Annual Review of Condensed Matter Physics **9**, 329 (2018).
[11] G. Xu, H. Weng, Z. Wang, X. Dai, and Z. Fang, Physical Review Letters **107**, 186806 (2011).
[12] F. D. M. Haldane, Physical Review Letters **61**, 2015 (1988).
[13] X. Feng, Y. Feng, J. Wang, Y. Ou, Z. Hao, C. Liu, Z. Zhang, L. Zhang, C. Lin, J. Liao, Y. Li, L.-L. Wang, S.-H. Ji, X. Chen, X. Ma, S.-C. Zhang, Y. Wang, K. He, and Q.-K. Xue, Advanced Materials **28**, 6386 (2016).
[14] M. M. Otrokov, I. I. Klimovskikh, H. Bentmann, A. Zeugner, Z. S. Aliev, S. Gass, A. U. B. Wolter, A. V. Koroleva, D. Estyunin, A. M. Shikin, M. Blanco-Rey, M. Hoffmann, A. Y. Vyazovskaya, S. V. Eremeev, Y. M. Koroteev, I. R. Amiraslanov, M. B. Babanly, N. T. Mamedov, N. A. Abdullayev, V. N. Zverev, B. Büchner, E. F. Schwier, S. Kumar, A. Kimura, L. Petaccia, G. Di Santo, R. C. Vidal, S. Schatz, K. Kißner, C.-H. Min, S. K. Moser, T. R. F. Peixoto, F. Reinert, A. Ernst, P. M. Echenique, A. Isaeva, and E. V. Chulkov, arXiv:1809:07389.
[15] D. Zhang, M. Shi, T. Zhu, D. Xing, H. Zhang, and J. Wang, Physical Review Letters **122**, 206401 (2019).
[16] J. Li, Y. Li, S. Du, Z. Wang, B.-L. Gu, S.-C. Zhang, K. He, W. Duan, and Y. Xu, Science Advances **5**, eaaw5685 (2019).
[17] Y. Gong, J. Guo, J. Li, K. Zhu, M. Liao, X. Liu, Q. Zhang, L. Gu, L. Tang, X. Feng, D. Zhang, W. Li, C. Song, L. Wang, P. Yu, X. Chen, Y. Wang, H. Yao, W. Duan, Y. Xu, S.-C. Zhang, X. Ma, Q.-K. Xue, and K. He, Chinese Physics Letters **36**, 076801 (2019).
[18] M. M. Otrokov, I. P. Rusinov, M. Blanco-Rey, M. Hoffmann, A. Y. Vyazovskaya, S. V. Eremeev, A. Ernst, P. M. Echenique, A. Arnau, and E. V. Chulkov, Physical Review Letters **122**, 107202 (2019).
[19] Y. Peng and Y. Xu, arXiv:1809:09112.
[20] J. Cui, M. Shi, H. Wang, F. Yu, T. Wu, X. Luo, J. Ying, and X. Chen, Physical Review B **99**, 155125 (2019).
[21] A. Zeugner, F. Nietschke, A. U. B. Wolter, S. Gaß, R. C. Vidal, T. R. F. Peixoto, D. Pohl, C. Damm, A. Lubk, R. Hentrich, S. K. Moser, C. Fornari, C. H. Min, S. Schatz, K. Kißner, M. Ünzelmann, M. Kaiser, F. Scaravaggi, B. Rellinghaus, K. Nielsch, C. Heß, B. Büchner, F. Reinert, H. Bentmann, O. Oeckler, T. Doert, M. Ruck, and A. Isaeva, arXiv:1812:03106.
[22] S. Huat Lee, Y. Zhu, Y. Wang, L. Miao, T. Pillsbury, S. Kempinger, D. Graf, N. Alem, C.-Z. Chang, N. Samarth, and Z. Mao, Physical Review Research **1**, 012011(R) (2019).
[23] Y. Deng, Y. Yu, M. Zhu Shi, J. Wang, X. H. Chen, and Y. Zhang, arXiv:1904.11468.
[24] C. Liu, Y. Wang, H. Li, Y. Wu, Y. Li, J. Li, K. He, Y. Xu, J. Zhang, and Y. Wang, arXiv:1905.00715.
[25] Z. S. Aliev, I. R. Amiraslanov, D. I. Nasonova, A. V. Shevelkov, N. A. Abdullayev, Z. A. Jahangirli, E. N. Orujlu, M. M. Otrokov, N. T. Mamedov, M. B. Babanly, and E. V. Chulkov, Journal of Alloys and Compounds **789**, 443 (2019).
[26] C. Hu, X. Zhou, P. Liu, J. Liu, P. Hao, E. Emmanouilidou, H. Sun, Y. Liu, H. Brawer, A. P. Ramirez, H. Cao, Q. Liu, D. Dessau, and N. Ni, arXiv:1905.02154.
[27] J. Wu, F. Liu, M. Sasase, K. Ienaga, Y. Obata, R. Yukawa, K. Horiba, H. Kumigashira, S. Okuma, T. Inoshita, and H. Hosono, arXiv:1905.02385.
[28] R. C. Vidal, A. Zeugner, J. I. Facio, R. Ray, M. H. Haghighi, A. U. B. Wolter, L. T. Corredor Bohorquez, F. Caglieris, S. Moser, T. Figgemeier, T. R. F. Peixoto, H. Babu Vasili, M. Valvidares, S. Jung, C. Cacho, A. Alfonsov, K. Mehlawat, V. Kataev, C. Hess, M. Richter, B. Büchner, J. van den Brink, M. Ruck, F. Reinert, H. Bentmann, and A. Isaeva, arXiv: 1906.08394

[29]     H. Sun, B. Xia, Z. Chen, Y. Zhang, P. Liu, Q. Yao, H. Tang, Y. Zhao, H. Xu, and Q. Liu, Physical Review Letters **123**, 096401 (2019).
[30]     D. Pacilè, S. V. Eremeev, M. Caputo, M. Pisarra, O. De Luca, I. Grimaldi, J. Fujii, Z. S. Aliev, M. B. Babanly, I. Vobornik, R. G. Agostino, A. Goldoni, E. V. Chulkov, and M. Papagno, physica status solidi (RRL) – Rapid Research Letters **12**, 1800341 (2018).
[31]     D. Souchay, M. Nentwig, D. Günther, S. Keilholz, J. de Boor, A. Zeugner, A. Isaeva, M. Ruck, A. U. B. Wolter, B. Büchner, and O. Oeckler, Journal of Materials Chemistry C **7**, 9939 (2019).
[32]     M. N. Baibich, J. M. Broto, A. Fert, F. N. Van Dau, F. Petroff, P. Etienne, G. Creuzet, A. Friederich, and J. Chazelas, Physical Review Letters **61**, 2472 (1988).
[33]     G. Binasch, P. Grünberg, F. Saurenbach, and W. Zinn, Physical Review B **39**, 4828 (1989).
[34]     J. Q. Yan, S. Okamoto, M. A. McGuire, A. F. May, R. J. McQueeney, and B. C. Sales, Physical Review B **100**, 104409 (2019).

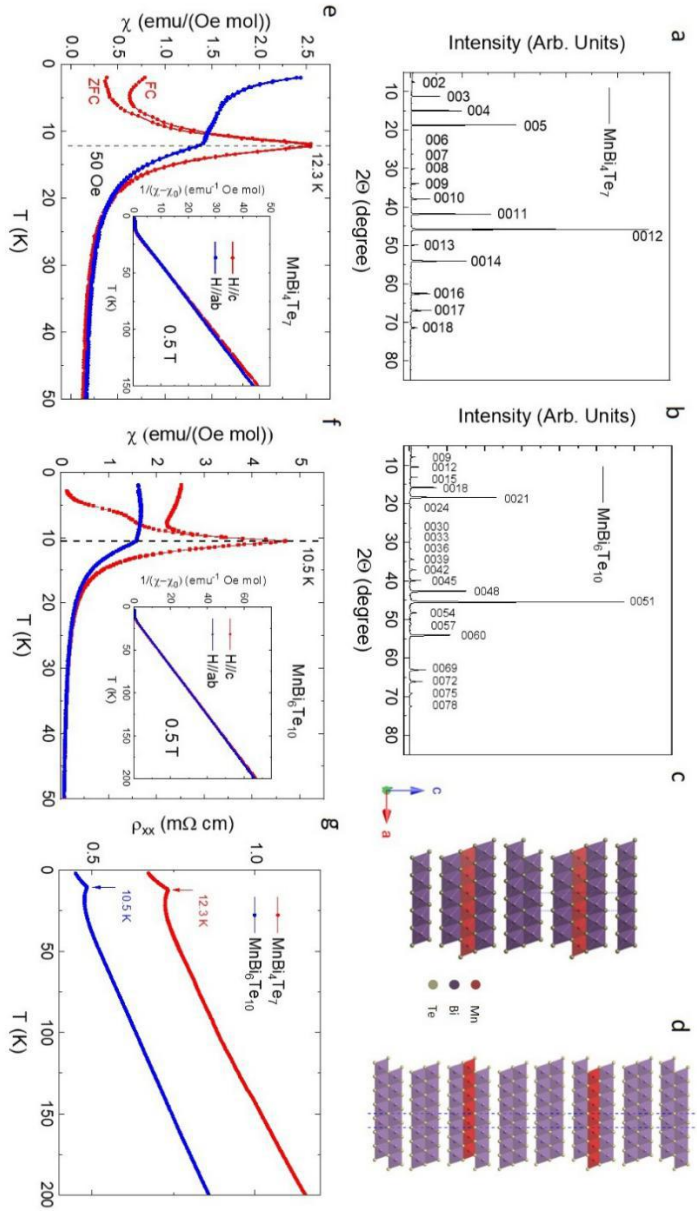

**Fig. 1, (a) and (b):** X-ray diffraction patterns with strong preferred orientation along [001] for $MnBi_2Te_4(Bi_2Te_3)_n$ (n=1,2); **(c) and (d):** Atomic structural models of $MnBi_2Te_4(Bi_2Te_3)_n$ (n=1,2). The purple blocks represent edge-sharing nonmagnetic $BiTe_6$ octahedra, and the red blocks indicate edge-sharing magnetic $MnTe_6$ octahedra. **(e) and (f):** Temperature dependence of zero field-cooled (ZFC) and field-cooled (FC) magnetic susceptibility for $MnBi_2Te_4(Bi_2Te_3)_n$ (n=1,2). The red and blue curves represent magnetic susceptibility with H//c and H//ab, respectively. The dashed lines indicate the antiferromagnetic transition temperature. The insets represent the temperature dependence of $1/(\chi - \chi_0)$ for FC curves. **(g):** In-plane electrical resistivity of $MnBi_2Te_4(Bi_2Te_3)_n$ (n=1,2) single crystals measured from room temperature down to 2 K. The red and blue arrows indicate the magnetic transition temperatures for $MnBi_2Te_4(Bi_2Te_3)_n$ (n=1,2).

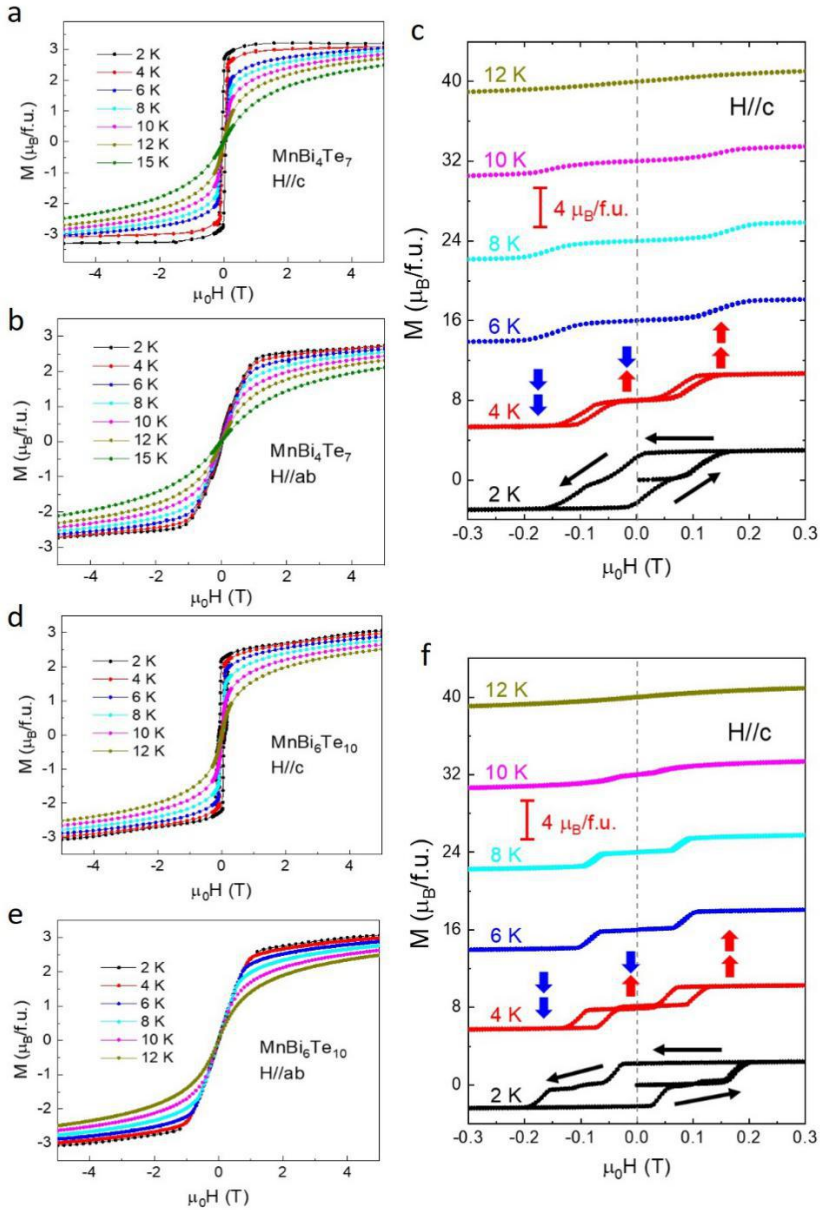

**Fig. 2**, Magnetic properties of MnBi$_2$Te$_4$(Bi$_2$Te$_3$)$_n$ (n=1,2) single crystals. **(a) and (b):** Isothermal magnetization for MnBi$_4$Te$_7$ with H//c and H//ab measured at various temperatures. **(d) and (e):** Isothermal magnetization for MnBi$_6$Te$_{10}$ with H//c and H//ab at various temperatures. **(c) and (f):** Enlarged area of M-H curves at low field for MnBi$_4$Te$_7$ and MnBi$_6$Te$_{10}$ with H//c. The M-H curves are shifted vertically with step of 8 $\mu_B$/f.u. for clarity. The sudden increase of M indicates the spin-flip transition. Ferromagnetic hysteresis can be observed at low temperatures with H//c. The red and blue arrows represent the spin orientations of the MnBi$_2$Te$_4$ SLs.

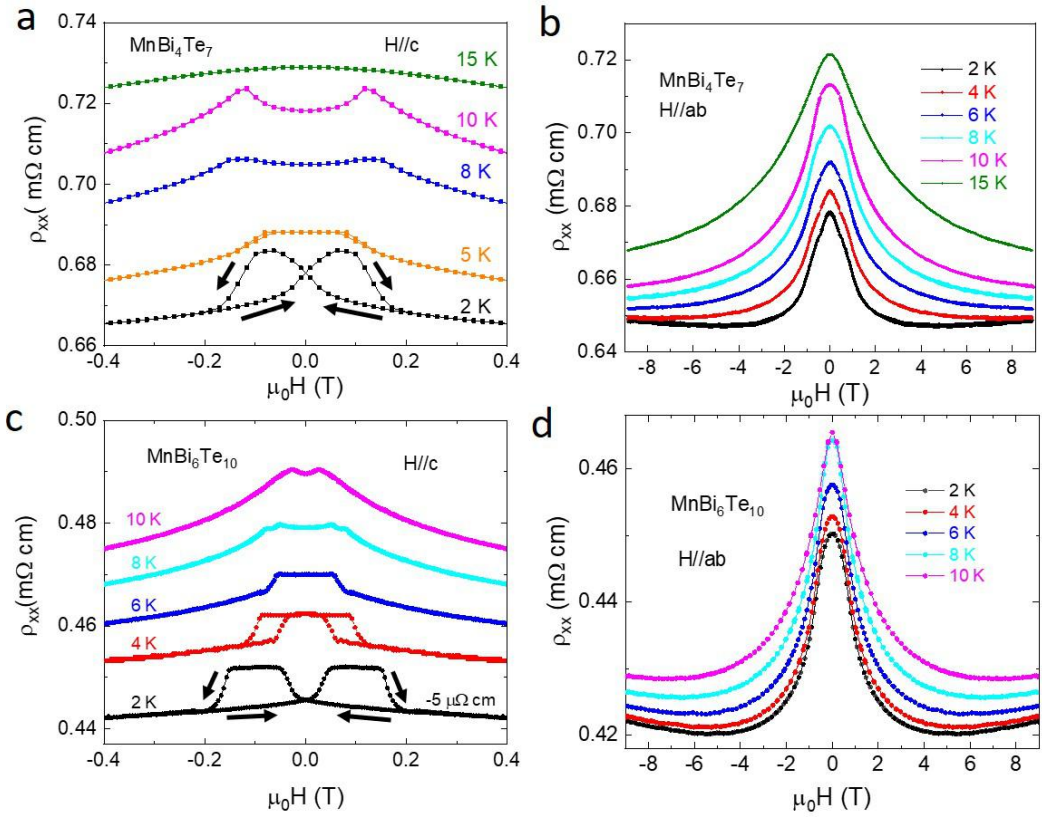

**Fig. 3,** Isothermal magnetoresistance of $MnBi_2Te_4(Bi_2Te_3)_n$ (n=1,2) single crystals at different temperatures. **(a) and (c):** Field dependence of in-plane resistivity $\rho_{xx}$ (H) with magnetic field applied parallel to the c axis for $MnBi_2Te_4(Bi_2Te_3)_n$ (n=1,2). The $\rho_{xx}$ (H) for $MnBi_6Te_{10}$ at 2 K is shift down by 5 $\mu\Omega$ cm for clarity. **(b) and (d):** Field dependence of in-plane resistivity $\rho_{xx}$ (H) with magnetic field applied in the ab-plane for $MnBi_2Te_4(Bi_2Te_3)_n$ (n=1,2) at various temperatures, respectively. The anomalies in the $\rho_{xx}$ (H) for $MnBi_2Te_4(Bi_2Te_3)_n$ with H//c indicate the spin flip transitions and the low-temperature butterfly shape indicates the ferromagnetic hysteresis.

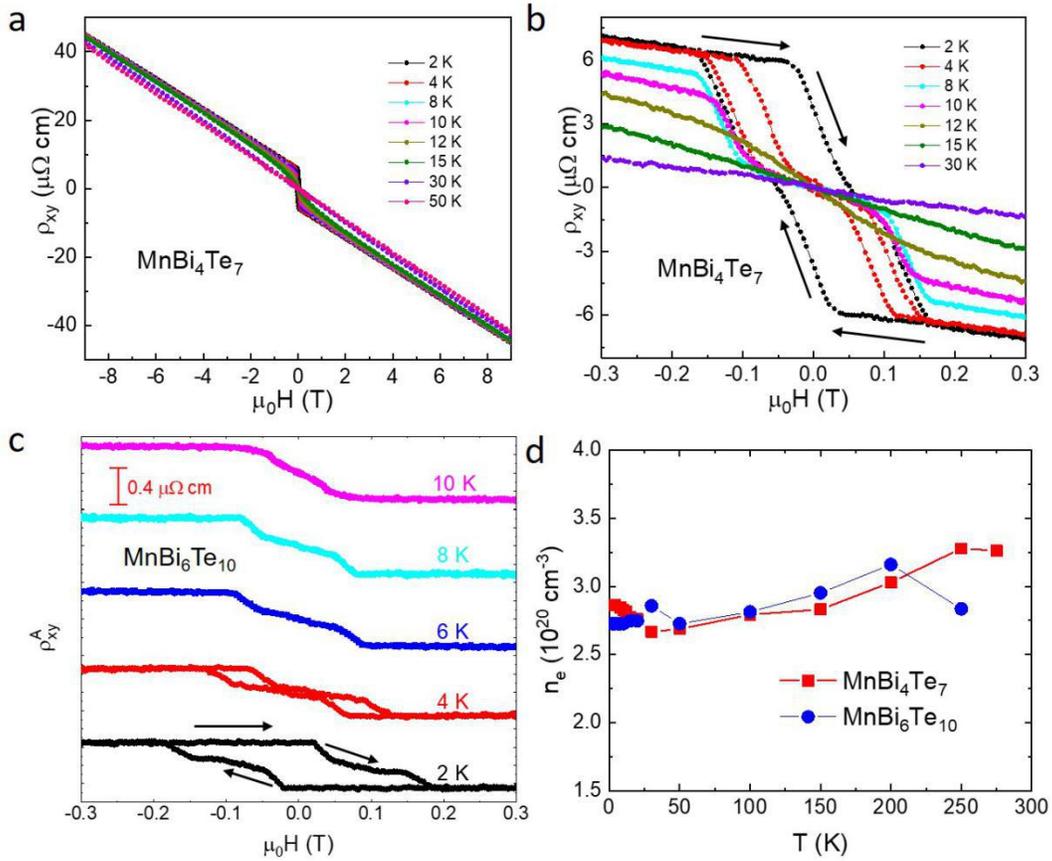

**Fig. 4, (a):** Hall resistivity ($\rho_{xy}$ (H)) as a function of magnetic field at various temperatures for MnBi$_4$Te$_7$ with magnetic field applied along the c-axis direction; **(b):** Enlarged regime of $\rho_{xy}$ (H) in the low magnetic field range from -0.3 to 0.3 Tesla clearly shows the anomalous Hall effect for MnBi$_4$Te$_7$; **(c):** Anomalous Hall resistivity ($\rho_{xy}^A$) as a function of magnetic field at various temperatures for MnBi$_6$Te$_{10}$. The $\rho_{xy}^A$ are shifted vertically with step of 0.8 μΩ cm for clarity. **(d):** Temperature dependence of carrier density n$_e$ for MnBi$_2$Te$_4$(Bi$_2$Te$_3$)$_n$ (n=1,2). Both of the samples show weak temperature dependence of the carrier density.